\documentclass[9pt,conference]{IEEEtran}

\usepackage{algorithm}
\usepackage{algpseudocode}
\usepackage{amsmath}
\usepackage{float}
\usepackage{flafter}
\usepackage{array}
\usepackage{multirow}
\usepackage{caption}
\usepackage{graphicx}
\usepackage{subcaption}
\usepackage[most]{tcolorbox}
\usepackage{makecell}
\usepackage{tabularx}
\usepackage{array}
\usepackage{listings}
\usepackage{xcolor}
\usepackage{booktabs} 

\usepackage[framemethod=TikZ]{mdframed}

\newmdenv[
  backgroundcolor=gray!5,
  linecolor=gray!50,
  linewidth=0.5pt,
  roundcorner=4pt,
  innerleftmargin=8pt,
  innerrightmargin=8pt,
  innertopmargin=6pt,
  innerbottommargin=6pt,
  skipabove=10pt,
  skipbelow=10pt,
  font=\ttfamily\fontsize{8}{8},
  align=left,
  nobreak=true
]{llmoutput}

\definecolor{plusgreen}{rgb}{0,0.5,0}
\definecolor{minusred}{rgb}{0.6,0,0}

\lstdefinestyle{diffstyle}{
  basicstyle=\ttfamily\fontsize{8}{11}\selectfont,
  escapeinside={(*@}{@*)},
  showstringspaces=false,
  frame=none,
  backgroundcolor=\color{white},
  moredelim=**[is][\color{plusgreen}]{@+}{@+},
  moredelim=**[is][\color{minusred}]{@-}{@-},
}

\AtBeginDocument{%
  \providecommand\BibTeX{{%
    \normalfont B\kern-0.5em{\scshape i\kern-0.25em b}\kern-0.8em\TeX}}}

\def\BibTeX{{\rm B\kern-.05em{\sc i\kern-.025em b}\kern-.08em
    T\kern-.1667em\lower.7ex\hbox{E}\kern-.125emX}}

\begin{document}

\title{BugGen: A Self‑Correcting Multi‑Agent LLM Pipeline for Realistic RTL Bug Synthesis}

\author{
\IEEEauthorblockN{Surya Jasper}
\IEEEauthorblockA{\textit{Texas A\&M University} \\
College Station, Texas, USA \\
suryajasper@tamu.edu}
\and
\IEEEauthorblockN{Minh Luu}
\IEEEauthorblockA{\textit{Infineon Technologies} \\
Hanoi, Vietnam \\
danhminh.luu@infineon.com}
\and
\IEEEauthorblockN{Evan Pan, Aakash Tyagi, Michael Quinn, Jiang Hu, David Houngninou}
\IEEEauthorblockA{\textit{Texas A\&M University} \\
College Station, Texas, USA \\
\{pan.evan,tyagi,m.d.quinn,jianghu,davidkebo\}@tamu.edu}
}

\maketitle

\begin{abstract}
Hardware complexity continues to strain verification resources, motivating the adoption of machine learning (ML) methods to improve debug efficiency. However, ML-assisted debugging critically depends on diverse and scalable bug datasets, which existing manual or automated bug insertion methods fail to reliably produce. We introduce BugGen, a first of its kind, fully autonomous, multi-agent pipeline leveraging Large Language Models (LLMs) to systematically generate, insert, and validate realistic functional bugs in RTL. BugGen partitions modules, selects mutation targets via a closed-loop agentic architecture, and employs iterative refinement and rollback mechanisms to ensure syntactic correctness and functional detectability. Evaluated across five OpenTitan IP blocks, BugGen produced 500 unique bugs with 94\% functional accuracy and achieved a throughput of 17.7 validated bugs per hour—over five times faster than typical manual expert insertion. Additionally, BugGen identified 104 previously undetected bugs in OpenTitan regressions, highlighting its utility in exposing verification coverage gaps. Compared against Certitude, BugGen demonstrated over twice the syntactic accuracy, deeper exposure of testbench blind spots, and more functionally meaningful and complex bug scenarios. Furthermore, when these BugGen-generated datasets were employed to train ML-based failure triage models, we achieved high classification accuracy (88.1\%–93.2\%) across different IP blocks, confirming the practical utility and realism of generated bugs. BugGen thus provides a scalable solution for generating high-quality bug datasets, significantly enhancing verification efficiency and ML-assisted debugging.
\end{abstract}




\section{Introduction}


Modern hardware systems continue to grow in complexity, often incorporating billions of transistors on a single chip. This increased complexity significantly expands the scope and difficulty of design verification tasks. 
Verification efforts already consume more than half of the total hardware development time \cite{verificationstats}, with this fraction steadily increasing each year. The heightened complexity also results in large volumes of test failures, particularly during early front-end development and initial volume regressions, placing severe demands on debugging resources to pinpoint root causes or effectively triage these failures. Addressing these challenges requires significant human expertise, computational resources, and time commitment. \cite{phoenix}. 

Machine Learning (ML)-based debugging techniques have recently shown promise in mitigating inefficiencies associated with manual debugging, but they require thousands of diverse and appropriately labeled bug scenarios to train adequately \cite{mldebugreview}. While legacy bugs from prior designs serve value, they tend to over-represent known failure modes and outdated design patterns, leading to overfitting and poor transferability. Thus, generating unique and realistic bug scenarios in contemporary design contexts is critically important. Manual bug insertion is intrinsically unscalable and limited in coverage \cite{encarsia}. Automated approaches, like constrained-random mutation, often generate trivial or implausible bugs that fail to simulate real engineer-induced errors \cite{buglocalization}. 


Furthermore, for inserted bugs to serve a meaningful role in ML-based debugging, they must not only be syntactically valid but also functionally detectable, meaning they are capable of triggering downstream failures under test regressions. Without this, they offer no insight for training debugging models. Determining this functional detectability requires re-running validation infrastructure, which is often gated by slow and resource-intensive simulation. This makes manual bug insertion heavily constrained by trial-and-error, while automated techniques that flood designs with obscure and unrealistic bugs needlessly consume simulation resources without offering meaningful and unique insights.
 
Consequently, generating realistic, diverse, and functionally detectable bugs at scale remains an unmet necessity. To address this, we present \textbf{BugGen}, a first of its kind, self-correcting, multi-agent LLM pipeline that autonomously generates, inserts, and validates realistic bugs in RTL. By utilizing a suite of modular LLM agents with shared memory and rollback, BugGen iteratively produces unique, syntactically valid, and functionally detectable bugs---mimicking the semantics and nuance of real-world mistakes.

Evaluated across five OpenTitan \cite{opentitan} designs, BugGen synthesized 500 bug scenarios with 94\% functional accuracy and an unattended throughput of 17.7 functionally meaningful bugs / hour. It also surfaced 104 undetected bugs in production regressions, highlighting gaps in test coverage. BugGen also outperforms Synopsys' \textit{Certitude} \cite{certitude}, producing more functionally meaningful and complex bug scenarios with twice the syntactic accuracy and overwhelmingly higher exposure of testbench blind spots.

\section{Previous Work}


Traditionally, engineers manually insert bugs into HDL code in order to simulate design flaws and compile large datasets for training ML-based debugging models.

First, engineers must decide where in the design to insert a bug and what type of bug to insert. These decisions are inherently biased and often result in engineers disproportionately injecting faults into regions of the design that they are most familiar with, where existing checks are already concentrated \cite{mutationtofault}. This can lead to blind spots in verification and uneven datasets that reduce model effectiveness on underrepresented areas.

Engineers must re-run the entire verification suite after each bug insertion, often requiring iterative refinement through trial and error. Consequently, manual insertion is tedious and impractical for modern large-scale designs.

As a result, the hardware verification community has turned to a range of automated approaches to introduce and detect bugs.

\subsection{Constrained-random Automation}

The most common automated bug insertion techniques rely on a constrained-random, mutation-based paradigm. Tools like \textit{Mantra} \cite{mantra} and Synopsys' \textit{Certitude} \cite{certitude} inject predefined syntactic changes, such as control path rewrites or operator inversions, to simulate simple, commonly observed bugs. They then automatically test the mutated design against the verification suite to determine if the bug is detectable.

Despite their automation, these mutation-based approaches remain fundamentally constrained in complexity, coverage, and adaptability. They rely on predefined syntactic templates without semantic awareness or contextual understanding of design behavior. This leads to datasets of trivial, redundant, and unrealistic bugs \cite{buglocalization}. This shortfall is particularly damaging when generating datasets for ML-based debug, as models trained on these overly simplistic examples will struggle to root-cause and triage the kinds of complex, functional bugs encountered in practice. Even for test suite validation, these mutation pipelines under-represent subtle corner cases and instead overpopulate the dataset with contrived, easily detectable errors that do not reflect genuine engineering mistakes \cite{encarsia}. 

\subsection{LLMs in Verification}
While LLMs have not been formally explored for automated bug generation, they have recently shown strong capabilities in code generation. Within this avenue, they have demonstrated several skills -- such as syntactic fluency, logical reasoning, and functional intuition -- that are equally valuable when applied to the inverse problem of bug insertion.

For instance, \textit{VeriGen} \cite{verigen} utilized LLMs to generate Verilog code from higher-level engineer specifications. While these efforts were geared toward correct code generation, they would frequently output erroneous blocks, which reveals a valuable insight — if properly guided, unconstrained LLMs can potentially be used to intentionally produce bugs that resemble the types of subtle mistakes humans would make.

Soon after, \textit{AutoChip} \cite{autochip} emerged, which, while still aimed at generating correct Verilog code, introduced a self-correcting loop to improve the success rate of generated code. This highlights the ability of agentic LLM pipelines to learn from and rectify their past mistakes to produce functionally useful outputs.

We propose that these insights from LLM-based RTL code generation systems can be applied to the inverse problem of intentional bug generation. If LLMs, when prompted in an agentic workflow, can reproduce human-like semantic and structural patterns, then with a specially designed mutation paradigm, they can just as effectively be used to produce realistic, functionally meaningful bugs. 



\section{Methods}

In this work, we present a novel solution to the problem of automated bug generation: an agentic, self-correcting, and learning-based pipeline that leverages LLMs to generate and validate realistic bugs in RTL modules. Our system is designed to be modular, design-agnostic, and fully autonomous. Additionally, the system is HDL language-agnostic. We frame the problem as a sequence of sequential, abstracted steps that are each solved by specialized agents connected in an interactive pipeline. Furthermore, each generated bug is automatically evaluated for syntactic correctness and functional detectability, and revised accordingly. Over time, the pipeline learns from past attempts and improves performance through in-context adaptation. 

The rest of this section is organized as follows: After a brief overview, we explain how RTL modules are intelligently partitioned into manageable segments (Module Splitter) to ensure efficient handling by LLMs. Subsequently, we describe our Class-based Mutation Paradigm, highlighting how predefined mutation types facilitate realistic bug scenarios. Next, we detail the automated multi-agent Mutation Pipeline, clarifying how specific mutations are selected, generated, validated, and refined in a closed-loop manner. Finally, we address the Parallelism strategy employed, demonstrating how BugGen’s architecture efficiently scales across multiple modules and designs.

\subsection{Overview}



For any given design, our bug generation pipeline works on a set of RTL files selected by the verification engineer. Each RTL file undergoes a multi-step, agentic workflow:

\begin{enumerate}
    \item Partition the module into a comprehensive set of regions
    \item Iteratively insert bugs into regions
    \item Evaluate the modified design to confirm the presence of unique, syntactically valid, and functionally detectable bugs
\end{enumerate}

Since directly processing entire large RTL modules with LLMs can exceed context limitations and introduce biases, we first introduce a specialized module-splitting stage to logically partition modules into manageable segments.

\subsection{Module Splitter}

\subsubsection*{Motivation \& Mutation Target Regions}

In SystemVerilog, a \textbf{module} is a block of code that represents a functional unit of the design. Modern SoCs contain hundreds of such modules, many of which span thousands of lines of deeply nested logic. Generating bugs in such modules using LLMs presents several challenges, including limited context windows, difficulty handling intricate code structures, and inconsistent outputs across runs. These constraints make it impractical to process entire modules directly without some form of abstraction.


To address these constraints and ensure an even spread of bugs, we divide each module into \textbf{mutation target regions} (MTRs) -- logically discrete segments of modules that encapsulate functionally meaningful behavior. These include finite state machines, control logic, or memory-mapped interfaces, where injected bugs are more likely to influence architectural outputs. The system gives engineers the choice to explicitly specify MTRs, in which case the pipeline uses only these regions for mutation. Otherwise, the pipeline utilizes the \textbf{module splitter}, a specialized agent designed to automatically partition modules into MTRs.



\subsubsection*{Automated Partitioning}

For each module, the module splitter is provided with the RTL code and a set of general guidelines to help define regions without breaking apart logically or functionally dependent code.

If the module is small enough to fit entirely within the LLM’s context window, the splitter processes the full module at once and outputs a complete partition of regions.

Larger modules require an iterative, context-aware approach. At each iteration, the splitter receives a fixed-size chunk and 5 auxiliary lines from the next chunk. The agent, guided by prompt instructions, splits the chunk into regions and determines whether the auxiliary lines logically continue the last region. This is done by instructing the agent to analyze whether the final region’s structure — such as a module, block, or expression — appears incomplete on its own based on Verilog syntax patterns, but would otherwise be complete with the auxiliary lines. If a dependency is detected, the agent omits the final region so it can be fully reprocessed in the next chunk. Otherwise, it includes all the regions and continues from the end of the chunk.

For each region, the splitter is also instructed to generate a brief description of the region’s functional purpose, which we will term the \textit{region synopsis} for clarity. This process repeats until the splitter pipeline produces a comprehensive module partition of regions, each with a description of its function and purpose for mutations to be inserted later. 

With modules now effectively partitioned into meaningful regions, the next critical step is to define a controlled mutation strategy which ensures that inserted bugs are realistic, meaningful, and syntactically valid. To this end, we introduce a class-based mutation paradigm.




\subsection{Class-based Mutation Paradigm}

\subsubsection*{Overview}

In our pipeline, a \textbf{mutation} refers to a syntactically valid modification applied to one or more consecutive lines of Verilog code---such as altering a conditional, modifying loop behavior, redirecting FSM transitions, etc. A \textbf{bug scenario} consists of multiple related mutations inserted simultaneously within a single module. These mutations may affect adjacent lines or span multiple regions, but they are functionally related and evaluated as a single unit. By grouping mutations into these bug scenarios, the pipeline can generate more sophisticated, realistic, and varied design flaws. The engineer can specify how many mutations should be included per bug scenario based on desired complexity.

As all LLMs have inherent stochasticity and knowledge deficits, it is crucial that we strike a balance between automation and control when producing these bug scenarios. Thus, we propose a multi-step, agentic, class-based mutation paradigm. This will allow verification engineers to have high-level control over the types of bugs that are inserted, while still providing enough automation to ensure varied and meaningful bug scenarios. We will now explore the steps involved in this class-based mutation process.

\subsubsection*{Mutation Index}

The \textbf{mutation index} provides a list of all the mutation classes that are available to the bug generator along with a brief description of each. This index is intentionally succinct, describing the functional aspects of each mutation and their desired end behaviors at a high-level context. We provide a baseline index of mutations, but it can easily be extended to include more complex mutation types without any additional granularity. For a more detailed overview, please refer to Appendix~\ref{appendix:mutationtypes}.

Each mutation class in the index will have a corresponding \textbf{mutation specification}, written by the verification engineer, which comprehensively details its function with syntactic examples. Verification engineers can choose to either use a single mutation index for all modules, or to provide multiple indexes customized for different designs and modules.


\subsubsection*{History and Cache}

To enable learning from past attempts, the pipeline maintains a shared \textbf{mutation cache} that records all previously attempted mutations along with their evaluation outcomes. This cache spans modules, designs, and even parallel threads to provide a persistent, design-agnostic memory that guides future mutation decisions. By referencing this cache, the pipeline avoids redundant mutations, prioritizes strategies with higher success rates, and improves over time by building a dynamic rulebook for syntactic correctness and functional utility.

Having clearly defined our mutation strategy, we now integrate this paradigm into a systematic, multi-agent mutation pipeline. This pipeline automates the region selection, mutation insertion, and validation processes, ensuring efficiency, diversity, and consistency of bug generation.

\subsection{Mutation Pipeline}

\subsubsection*{Overview}

The full mutation pipeline involves a multi-step process consisting of three independent LLM-driven agentic steps, followed by evaluation.

\begin{figure}[h]
    \centering
    \includegraphics[width=\linewidth]{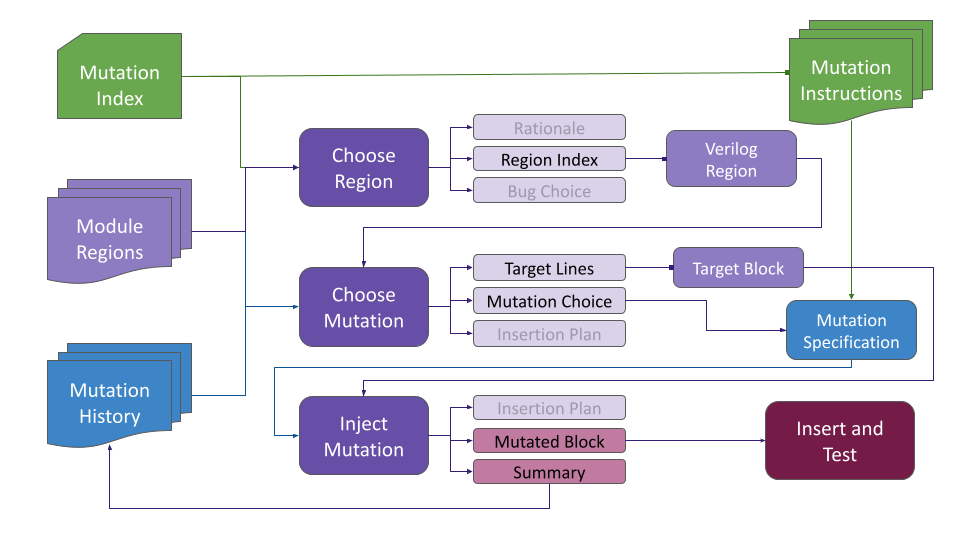}
    \caption{LLM Mutation Pipeline}
    \vspace{-12pt}
    \label{fig:mutationpipeline}
\end{figure}

A visual diagram illustrating this process is shown in \textit{Figure~\ref{fig:mutationpipeline}}. The agentic components will be detailed in the following subsections. For a more detailed breakdown of these steps with examples of outputs generated by each LLM agent, please refer to \textit{Appendix~\ref{appendix:llmexamples}}.

\subsubsection*{Step 1: Select Region}
The \textbf{region selector agent} chooses a region from the module partition based on three primary criteria: (1) surface coverage, prioritizing regions that have seen fewer previous mutations to ensure broad distribution of inserted bugs; (2) success rates, favoring regions with historically higher probabilities of generating syntactically valid and functionally detectable bugs; and (3) uniqueness, emphasizing selection of regions likely to yield mutations distinct from those already explored, thereby increasing the diversity of the generated dataset.

To balance these criteria, we provide the agent with the following:

\begin{enumerate}
    \item Module partition: includes the region synopsis of all available regions in the module along with the number of mutations that have previously been inserted into them. The agent is instructed to select regions that are likely to trigger interesting end-behavior, and have fewer attempted mutations to ensure broader coverage.
    \item Mutation attempt history: includes the overall success rate of mutations and distribution of mutation classes within each region. The agent is instructed to avoid regions that consistently produce undetectable bugs. Note that this feature may be disabled if the system is being used to assess verification infrastructure robustness.
    \item Mutation index: The agent uses this in conjunction with the mutation attempt history to prioritize regions where underrepresented mutation classes are likely applicable.
\end{enumerate}

With these inputs, the region selector chooses a region and also outputs a rationale and proposed mutation class. The latter are not used directly, but are included to encourage deliberate reasoning about future mutation success.

\subsubsection*{Step 2: Select Mutation}

With a region selected, the \textbf{mutation selector agent} then chooses the specific mutation to inject. It is provided with the following:

\begin{enumerate}
    \item Mutation index, specifying all allowed mutation classes
    \item RTL code of the selected region
    \item Region-specific mutation history, organized by whether each past attempt succeeded to produce a detectable bug
\end{enumerate}

These inputs allow the agent to avoid past mistakes, promote variety, and prioritize insertion strategies that yield distinct and realistic bug scenarios, closely resembling mistakes a human designer could likely introduce.

Given this context, the agent selects both a mutation class and the \textbf{target block} for insertion. For single-line mutations, it selects a single line for the target block; for multi-line mutations, it may choose anywhere from 1 to 4 lines. We also prompt it to produce a tentative insertion plan to promote thoughtful reasoning.

\subsubsection*{Step 3: Inject Mutation}

With the mutation selected, the pipeline proceeds with the \textbf{mutation injector agent}, which generates and inserts the mutation syntactically. First, the pipeline extracts the detailed mutation specification based on the selected mutation class and provides it to the agent.

We provide the selected target block along with the full RTL code of the region so that the injector can utilize the surrounding context.

With this information, the agent mutates the target block alone according to the mutation specification. This mutated block is then injected into the design. We require the agent to provide a mutation summary, describing the functional purpose of the inserted mutation.

The mutated block and summary are combined into a new mutation entry, which is added to the mutation history. Upon evaluation, this entry's success tag will be updated depending on whether it produced a detectable bug or not.

\subsubsection*{Step 4: Evaluate}

Steps 1--3 of the mutation pipeline are repeated for as many mutations as requested per \textbf{bug scenario}. 


The pipeline then verifies structural uniqueness by comparing each mutation against the shared mutation cache. If any of them are redundant, the pipeline moves back to step 1.

Next, the pipeline attempts to compile the design. If compilation fails, this reveals that at least one of the mutations was syntactically invalid. Thus, the corresponding mutation entries for the given bug scenario are labeled as failed attempts within the mutation history. If compilation succeeds, the pipeline proceeds with the simulator for functional evaluation.

The test regression suite we use in our evaluation consists entirely of pre-existing test cases written by the OpenTitan team. These test cases are designed to verify functional correctness across various scenarios within the hardware design. To expand coverage, the full suite is rerun with different random seeds, which effectively creates new variations of the test cases and enables a broader evaluation of the design's behavior across different input cases.

If an injected bug scenario succeeds in compilation and triggers at least one test case failure, the mutation entries are labeled as successful as they resulted in a genuine deviation from expected behavior. Otherwise, the mutation entries are labeled as failures since they either failed to compile or failed to induce a detectable functional bug. Regardless of the evaluated success metric, the attempted mutations are added to the design-agnostic, shared mutation cache. This allows parallel threads and future runs to evolve over time.

\subsubsection*{Step 5: Repeat}

Lastly, the pipeline cycles back to step 1 with the bug-free design and updated history. If the evaluation was successful, it will proceed with a new bug scenario. Otherwise, it will roll back the mutations from the unsuccessful iteration and retry the current bug scenario.

With the methodology fully detailed including partitioning, mutation definition, pipeline execution, and parallelization, we proceed in the next section to experimentally validate our approach, measuring its performance against key metrics across diverse benchmark designs.

\section{Experimental Setup}

To evaluate the performance of our pipeline, we run it across 5 designs within the OpenTitan open-source hardware framework. For each of these 5 designs, the pipeline is configured with a subset of SystemVerilog modules. These modules were selected for their structural variety and represent different functional domains within the larger system on chip. 

To enable a controlled comparison between Certitude and BugGen, we selected a custom MESI design, as Certitude's simulation configurations posed practical challenges when applied directly to the designs in OpenTitan open-source hardware framework. All designs are reflected in \textit{Table~\ref{tab:designmodules}} along with their respective sizes.

\begin{table}[h]
\centering
\footnotesize
\begin{tabular}{|l|l|c|c|}
\hline
\textbf{Design Group} & \textbf{Design} & \textbf{Modules} & \textbf{Total Lines} \\
\hline
\multirow{5}{*}{OpenTitan} 
 & AES Cipher Control & 7 & 2038 \\
 & I2C Bus Monitor     & 5 & 2927 \\
 & OTP Control         & 6 & 2771 \\
 & ROM Control         & 5 & 893 \\
 & USB Control         & 8 & 2821 \\
\hline
\multicolumn{1}{|l|}{Comparison} & MESI & 2 & 279 \\
\hline
\end{tabular}
\caption{Summary of Modules and Size per Design}
\vspace{-12pt}
\label{tab:designmodules}
\end{table}

For all LLM-driven components of the pipeline, we use the GPT-4o Mini model, which we found to strike the best balance between speed, consistency, and output quality. In preliminary experiments, we compared several open-weight models (including LLaMA 2/3 variants) and proprietary APIs (such as Gemini 2.0), and observed that GPT-4o Mini consistently delivered stable and syntactically valid completions at significantly lower latency.

Furthermore, we use a configuration of two mutations per bug scenario to strike a balance between bug scenario complexity and evaluation efficiency. The pipeline is initialized with the baseline mutation index and an empty mutation cache, meaning it is completely stranger to the provided designs at the beginning of the run. Thus, all in-context learning is conducted and measured within the scope of the run alone. Lastly, no mutation target regions are provided to the pipeline, so the module splitter is in play across all designs.

\section{Results}

We evaluate the quality and efficiency of generated bugs along six critical dimensions: \textbf{accuracy}, \textbf{autonomy}, \textbf{efficiency}, \textbf{coverage}, \textbf{adaptability}, and \textbf{extensibility}. The results below are drawn from the comprehensive set of 500 unique, detectable bug scenarios (1000 mutations total) generated, validated, and cached by the parallelized run across all 5 OpenTitan designs. 

\subsection{Accuracy}

\subsubsection*{Functional Accuracy}

We define \textbf{functional accuracy} as the successful creation of a bug scenario that is unique, syntactically valid, and functionally detectable. This metric is valuable for training AI-based debug models as they are dependent on the state of existing testing infrastructure and require examples that induce detectable downstream faults. 

Across all 5 designs, the pipeline achieved an overall functional accuracy of \textbf{94.2\%} within a maximum of two retries. This means that nearly all generated bug scenarios ultimately pass both syntactic and functional validation. Additionally, the pipeline demonstrates a first-attempt functional accuracy of \textbf{56.4\%}, indicating that a majority of bug scenarios are generated successfully without any revision or rollback. These average accuracies are presented alongside accuracy results by mutation class in Figure 2. 

\begin{figure}[h]
    \centering
    \includegraphics[width=\linewidth]{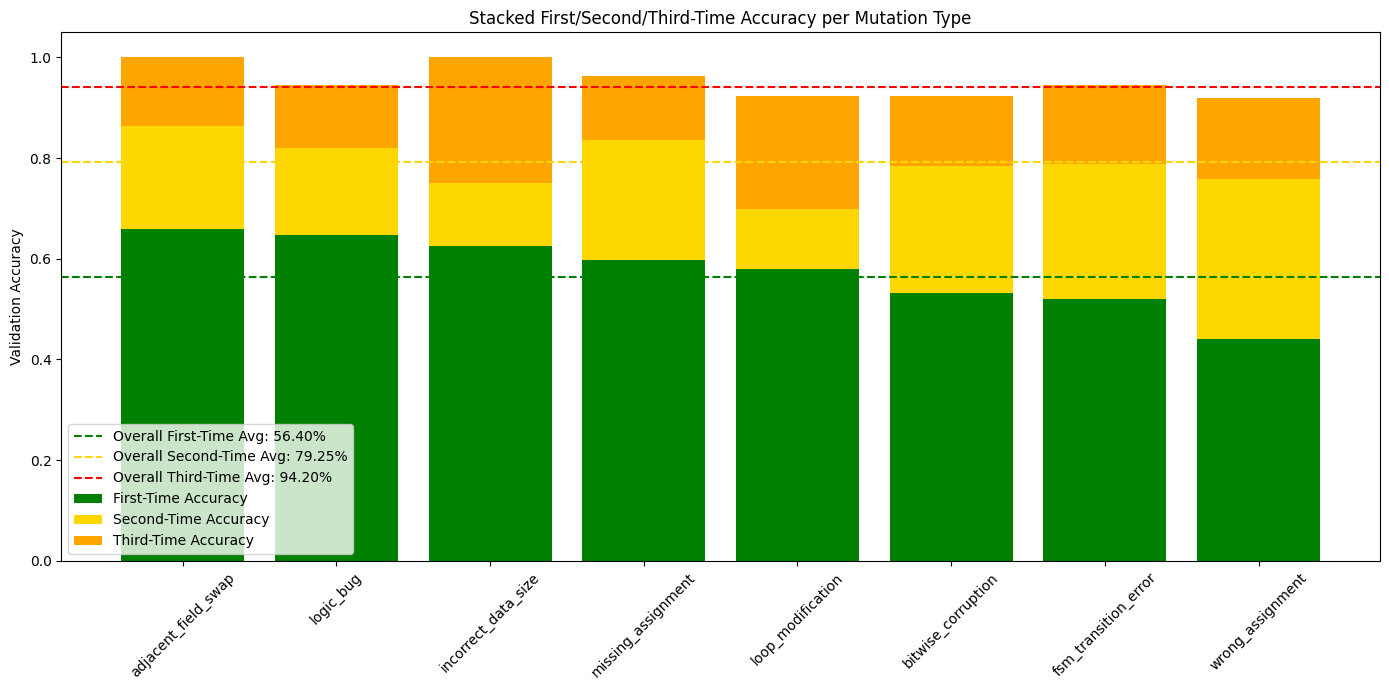}
    \caption{Validation Accuracy by Mutation Type}
    \vspace{-12pt}
    \label{fig:accuracy}
\end{figure}

\subsubsection*{Syntactic Accuracy}

On the other hand, in contexts where the pipeline is used to stress test verification suites or evaluate verification coverage, functionally undetected bugs are no longer a failure, but rather an indicator of blind spots in the test bench. For this, we simply evaluate \textbf{syntactic accuracy}, as any syntactically valid bugs would be useful in gauging test coverage. 

\textit{Table~\ref{tab:failurebreakdown}} details the first-attempt mutation results across all inserted mutations, showcasing a syntactic accuracy of approximately \textbf{64\%}. From this breakdown, we observe that 104 mutations (roughly 10\% of all inserted bugs) were syntactically valid yet undetected by the OpenTitan test volume regression suite. Furthermore, specific designs, such as the AES Cipher Control, account for a disproportionately higher number of these, revealing verification blind spots. 

\begin{table}[h]
\centering
\footnotesize
\begin{tabular}{|l|c|c|c|}
\hline
\textbf{OpenTitan Design} & \textbf{Detected} & \textbf{Syntax Failure} & \textbf{Undetected} \\
\hline
AES Cipher Control & 81 & 64 & 72 \\
I2C Bus Monitor & 85 & 44 & 0 \\
OTP Control & 110 & 129 & 14 \\
ROM Control & 145 & 62 & 18 \\
USB Control & 117 & 59 & 0 \\
\hline
\textbf{Total} & \textbf{538} & \textbf{358} & \textbf{104} \\
\hline
\end{tabular}
\caption{Bug Outcomes on OpenTitan Designs}
\vspace{-6pt}
\label{tab:failurebreakdown}
\end{table}

Furthermore, when evaluated against Certitude on the MESI modules, BugGen produced 36 undetected bugs compared to just 2 from Certitude, revealing significantly more testbench blind spots. Additionally, BugGen demonstrated over twice the syntactic accuracy. These results are summarized in Table~\ref{tab:outcome_comparison}.

\begin{table}[h]
\centering
\footnotesize
\begin{tabular}{|l|c|c|c|}
\hline
\textbf{MESI Module} & \textbf{Detected} & \textbf{Syntax Failure} & \textbf{Undetected} \\
\hline
LRS Arbiter & (75, 125) & (12, 51) & (11, 2) \\
LRU Block   & (58, 42)  & (17, 10) & (25, 0) \\
\hline
\textbf{Total} & (133, 167) & (29, 61) & (36, 2) \\
\hline
\end{tabular}
\caption{Bug outcomes across MESI modules comparing our pipeline (first value) with Certitude (second value)}
\vspace{-12pt}
\label{tab:outcome_comparison}
\end{table}

\subsection{Autonomy}

Our pipeline functions independently once initialized without human intervention at any stage. This is achieved in practice through a closed-loop interaction between LLM agents and validators. 

Each component of the pipeline – module partitioning, mutation insertion, evaluation, caching, and rollback – is fully automated. Once a mutation is attempted, the pipeline automatically checks for syntactic correctness through compilation and functional detectability through simulation. If a mutation fails at any stage, the pipeline automatically rolls back and retries until it succeeds.


\subsection{Efficiency}

In terms of raw throughput, the pipeline demonstrates high efficiency across the OpenTitan designs. The bug generation (LLM portion) takes approximately \textbf{16 seconds per bug scenario}, including rollback and retry logic. Validation (simulation and detection) time varies based on design size, averaging \textbf{6.4 minutes per bug scenario}. 

\begin{figure}[h]
    \centering
    \includegraphics[width=\linewidth]{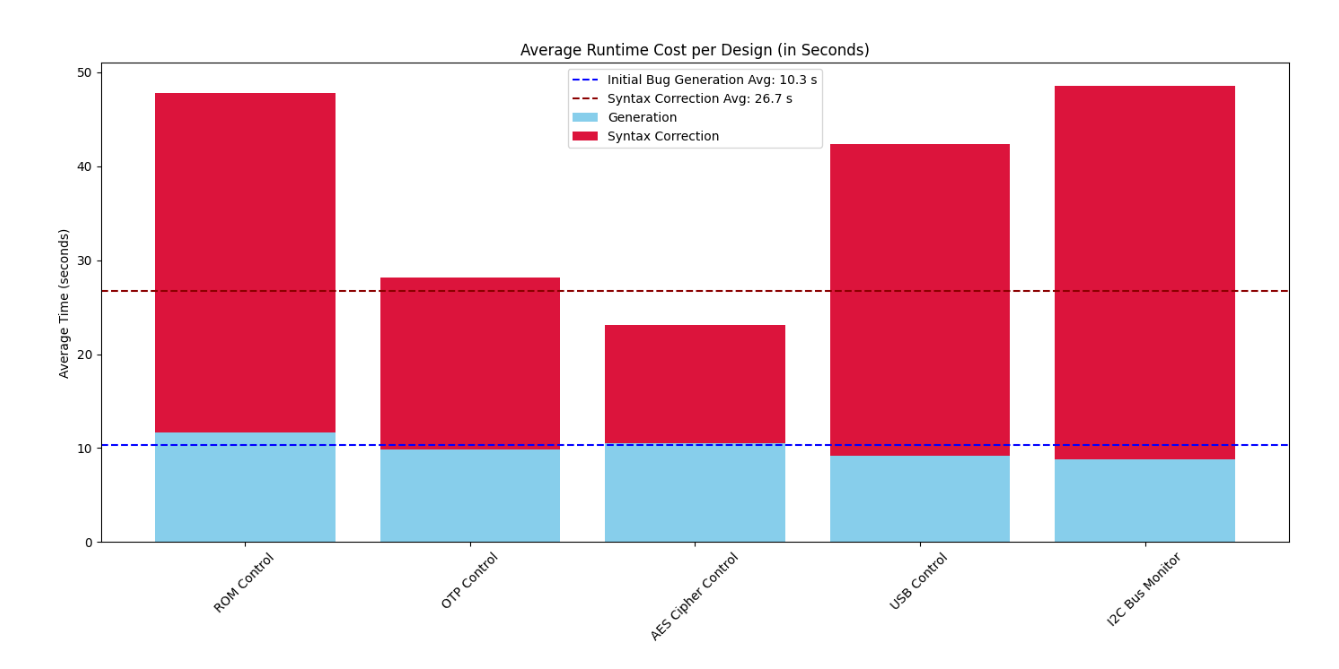}
    \caption{Correctable Runtime Cost by Design}
    \vspace{-5pt}
    \label{fig:runtimecost}
\end{figure}

To better isolate the cost of our correctable overhead, we extract the runtimes for initial bug generation and syntax correction. \textit{Figure~\ref{fig:runtimecost}} illustrates the average time spent per design in each category. In this case, the initial bug generation time captures the pipeline’s generation of the initial mutation, while the syntax correction time accounts for compilation, rollback, and re-generating mutations after failure. 

As shown in \textit{Figure~\ref{fig:runtimecost}}, across all designs, the average first-attempt generation time is 10.3 seconds, and the average syntax correction time is 26.7 seconds. This demonstrates that outside of the simulation time, our correctable overhead is relatively low.


\begin{table}[h]
\centering
\footnotesize
\begin{tabular}{|l|p{1.7cm}|p{1.4cm}|p{1.5cm}|}
\hline
\textbf{Design} & \textbf{Validation Time (minutes)} & \textbf{Total Time (minutes)} & \textbf{Bugs / Hour} \\
\hline
AES Cipher Control & 6.556 & 18.098 & 3.315 \\
I2C Bus Monitor & 21.849 & 35.011 & 1.714 \\
OTP Control & 4.158 & 8.379 & 7.161 \\
ROM Control & 5.751 & 7.795 & 7.697 \\
USB Control & 15.566 & 23.022 & 2.606 \\
\hline
\textbf{Overall Average} & 8.405 & 14.674 & 4.089 \\
\textbf{Parallel Execution} & 21.849 & 1,960.607 & \textbf{17.719} \\
\hline
\end{tabular}
\caption{Speed of Bug Insertion \& Validation}
\vspace{-12pt}
\label{tab:speed}
\end{table}

\textit{Table~\ref{tab:speed}} outlines the sequential performance metrics across each design thread along with the measured parallelized runtimes. This presents the key metric of \textbf{17.72 unique, detectable, and meaningful bug scenarios per hour} – significantly faster than any expert verification engineer. Moreover, this metric can be scaled even further with larger collections of designs, or intra-design parallelism. This demonstrates the critical advantage of this pipeline: theoretically unlimited parallelism restricted solely by simulation resources and the number of independent designs or RTL modules of interest. 

\subsection{Coverage}

We define coverage as a holistic metric that can be divided into 2 main categories: uniqueness and spread. We will demonstrate the pipeline’s success in each individually.

\textbf{Uniqueness} refers to the generation of structurally unique bugs. This is guaranteed by design as every produced mutation is automatically compared against the mutation cache during the evaluation stage to ensure there is no repetition. 



\textbf{Spread} refers to an even distribution of mutations across functionally meaningful sections of the design. We evaluate this by utilizing the manually segmented mutation target regions (MTRs) for each target RTL module. As these MTRs correspond to regions that human engineers have specifically tagged as functionally meaningful, we can use them in our evaluation of spread.

It is important to note that this experimental run deferred to the module splitter, so the pipeline never used or even looked at the engineer-provided MTRs. A high percentage of mutations within MTRs and an even distribution across these regions strongly indicate effective bug spread by our pipeline.

To compute the spread score, we first calculate the relative density of mutations in each MTR as:

$$
\text{relative\_mtr}_i = \frac{\text{\# mutations in MTR}_i}{\text{\# lines in MTR}_i}
$$

Then, we normalize this vector to form a probability distribution $p = [p_1, p_2, \ldots, p_N]$ over the $N$ MTRs, where:

$$
p_i = \frac{\text{relative\_mtr}_i}{\sum_{j=1}^N \text{relative\_mtr}_j}
$$

Lastly, we compute the normalized entropy of this distribution:

$$
H(p) = -\sum_{i=1}^{N} p_i \log p_i \,\Big/\, \log N
$$

This yields a score between 0 and 1, where 0 corresponds to complete concentration in one MTR and 1 corresponds to a perfectly uniform spread across all MTRs.

We also evaluate the percent of mutations that fall within the provided MTRs. The results of both metrics are reported in \textit{Table~\ref{tab:spread}}, showcasing a high degree of evenness in spread and near perfect selectivity of functionally meaningful regions, closely replicating the intuition of a verification engineer.


\begin{table}[h]
\centering
\footnotesize
\begin{tabular}{|l|c|c|}
\hline
\textbf{OpenTitan Design} & \textbf{Spread Score} & \textbf{Mutations in MTRs} \\
\hline
AES Cipher Control & 0.722124 & 92.23\% \\
I2C Bus Monitor & 0.799227 & 98.21\% \\
ROM Control & 0.671986 & 96.84\% \\
OTP Control & 0.702983 & 92.09\% \\
USB Control & 0.732082 & 100.00\% \\
\hline
\textbf{Average} & \textbf{0.725680} & \textbf{95.87\%} \\
\hline
\end{tabular}
\caption{Spread of Inserted Bugs in OpenTitan Designs}
\vspace{-4pt}
\label{tab:spread}
\end{table}

On the MESI modules, BugGen outperforms Certitude in both spread and MTR targeting, as shown in \textit{Table~\ref{tab:spread_comparison}}. This more strategic placement of mutations yields functionally richer bugs, resulting in more valuable bug datasets and improved exposure of testbench blind spots, as reflected earlier in \textit{Table~\ref{tab:outcome_comparison}}.

\begin{table}[h]
\centering
\footnotesize
\begin{tabular}{|l|c|c|}
\hline
\textbf{MESI Module} & \textbf{Spread Score} & \textbf{Mutations in MTRs} \\
\hline
ARB & (0.9011, 0.7742) & (89.34\%, 32.97\%) \\
LRU & (0.7282, 0.6428) & (99.50\%, 48.08\%) \\
\hline
\textbf{Average} & (\textbf{0.8146}, 0.7085) & (\textbf{94.42\%}, 40.52\%) \\
\hline
\end{tabular}
\caption{Spread of bugs in MESI modules, comparing our pipeline (first value) with Certitude (second value)}
\vspace{-9pt}
\label{tab:spread_comparison}
\end{table}

\subsection{Adaptability}

A key advantage of the pipeline is its ability to learn from past mistakes and adapt dynamically to unfamiliar designs through in-context learning based on the shared mutation cache. The system autonomously improves mutation accuracy over time by learning from past examples, without requiring manual tuning or explicit instructions.

\begin{figure}[h]
    \centering
    \includegraphics[width=\linewidth]{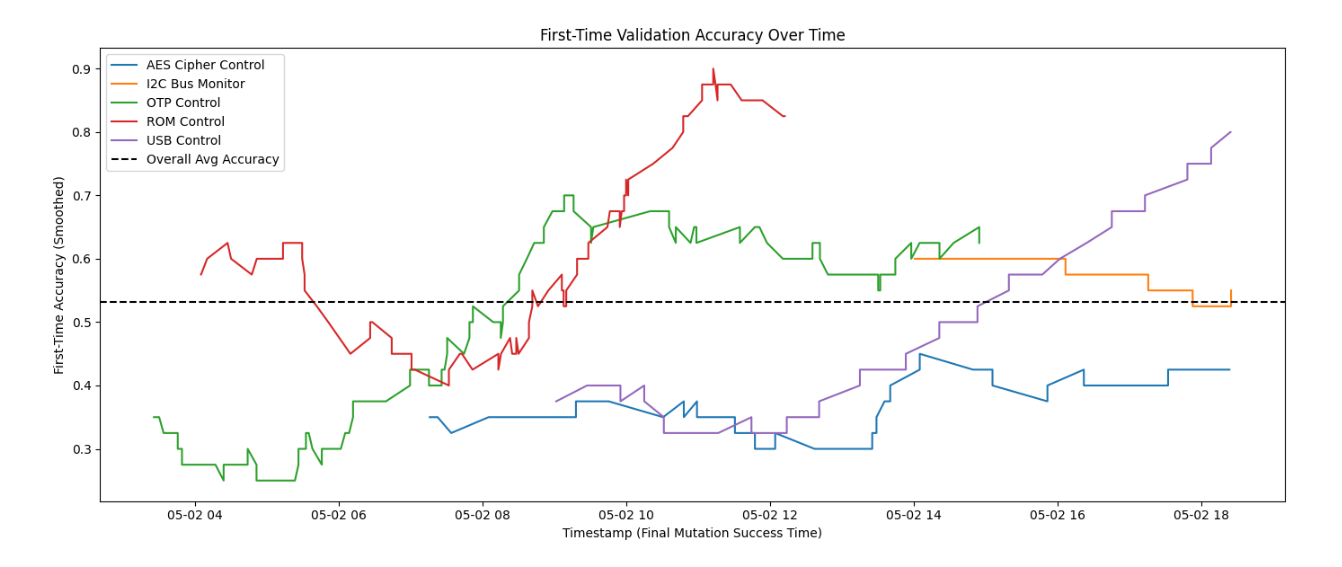}
    \caption{Evolution of Validation Accuracy Across Designs}
    \vspace{-12pt}
    \label{fig:accuracyevolution}
\end{figure}

This effect is shown in \textit{Figure~\ref{fig:accuracyevolution}}, where first-time mutation accuracy improves steadily across all five designs. This upward trend illustrates that as the shared cache grows, the pipeline successfully applies knowledge across design boundaries to reduce failures and generalize.

\subsection{Scalability}

Our pipeline is designed to scale seamlessly to large industrial designs. This is achieved through (1) the \textbf{module splitter}, which partitions large RTL files into logically distinct, workable regions; and (2) the pipeline’s support for both \textit{inter-design} and \textit{intra-design} parallelism (see Appendix~\ref{appendix:parallelism} for more detail). These features enable efficient operation on complex SoC designs, limited only by available simulation resources.




\subsection{Downstream Evaluation with ML-based Triage}
Lastly, to holistically evaluate the practical utility of the generated bug scenarios, we leveraged them to train a suite of machine learning models for failure triage using a ML-based triage tool built independently. 
These models were trained on the labeled failure waveform data corresponding to the validated bug scenarios produced by our pipeline. This provided them with a diverse and realistic training set, with between 80 - 120 unique bugs for each of 4 OpenTitan designs.

Table~\ref{tab:vcdiag} highlights the high failure classification accuracy across all four of the OpenTitan designs. 
This serves as a strong practical indicator of the realism and functional utility of our generated bugs as well as the overall design-agnostic capabilities of the pipeline. 

\begin{table}[h]
  \centering
  \footnotesize
  \setlength{\tabcolsep}{10pt}
  \renewcommand{\arraystretch}{1.2}
  \begin{tabular}{|c|c|c|c|c|}
    \hline
    \textbf{Design} & \textbf{Accuracy (\%)} & \textbf{Top-3 (\%)} &  \textbf{AUC (\%)} \\
    \hline
    AES      & 88.1  & 96.1  & 98.1  \\
    OTPCtrl  & 89.8  & 95.8  & 98.6  \\
    ROMCtrl  & 92.8  & 98.5  & 99.3  \\
    USB      & 93.2  & 97.9  & 99.8  \\
    \hline
  \end{tabular}
  \caption{Triage Accuracy with BugGen}
  \vspace{-6pt}
  \label{tab:vcdiag}
\end{table}

These results highlight BugGen’s effectiveness both as a verification stress-test tool and as a generator of high-quality datasets for ML-based debugging.

\section{Conclusion} 
Through a self-correcting, agentic architecture, BugGen demonstrates that LLMs can be used to autonomously generate realistic, functionally detectable RTL bug scenarios at scale. With over 500 unique bug scenarios synthesized across five diverse OpenTitan designs, our system showed strong performance across the key metrics of accuracy, efficiency, adaptability, and extensibility -- producing datasets that are well-suited for training ML models for root-cause analysis and failure prediction. Compared to existing industrial tools like Certitude, BugGen achieves higher syntactic accuracy, exposes significantly more testbench blind spots, and produces more functionally meaningful and structurally complex bug scenarios, all while requiring minimal manual effort.

While this work focused primarily on the generation of rich, scalable datasets for training ML-assisted debug, the implications extend further. Each unique, syntactically valid bug scenario that passes undetected through simulation offers a concrete signal of oversights in the testbench. This allows BugGen to function not only as a generator of training data but also as an analytical tool for evaluating and refining verification infrastructure. In the future, we aim to leverage this capability to automate testbench enhancement.


\bibliographystyle{IEEEtran}
\bibliography{buggen_sources}

\appendices

\section{Mutation Index}
Table~\ref{tab:mutationclasses} shows our full baseline mutation index used for evaluation. This can be easily extended by verification engineers to encompass more complex and specialized mutations. Each mutation can either be single-line or multi-line.

\label{appendix:mutationtypes}
\begin{table}[!h]
    \footnotesize
    \setlength{\tabcolsep}{10pt}
    \renewcommand{\arraystretch}{1.1}
    \begin{center}
      \begin{tabular}{|>{\centering\arraybackslash}m{2.1cm}|>{\raggedright\arraybackslash}p{4.8cm}|}
          \hline
          \textbf{Mutation Class} & \textbf{Mutation Description} \\
          \hline
          Missing assignment & When an assignment to a variable that should be there is missing. This can be achieved by commenting out an existing assignment statement. For example: \texttt{assign a = b;} $\rightarrow$ \texttt{// assign a = b;}. Do not introduce new variables. \\
          \hline
          Bitwise corruption & A bug that performs incorrect bitwise manipulation on the right-hand side of an assignment. For example: \texttt{a = b \& c;} $\rightarrow$ \texttt{a = b}. \\
          \hline
          Logic bug & A bug in the condition of an \texttt{if}, \texttt{while}, \texttt{for}, or \texttt{always} block (\texttt{always\_ff}, \texttt{always\_comb}) that results in erroneous behavior. Only applicable to conditional and always blocks. \\
          \hline
          Wrong assignment & A bug where the right-hand side of an assignment is correct, but it is assigned to the wrong variable. For example: \texttt{a = b;} $\rightarrow$ \texttt{c = b;}. Only use existing variables. \\
          \hline
          Incorrect data size & A bug where the array size is changed, leading to data mismatches or misalignment. \\
          \hline
          Adjacent field swap & Swaps fields between two adjacent, functionally-related assignments (LHS or RHS). Example: \texttt{a = b; c = d;} $\rightarrow$ \texttt{a = d; c = b;} \\
          \hline
          Loop modification & Alters loop start/end/increment or internal logic to change iteration behavior. This includes off-by-one and boundary bugs. Example: \texttt{for (i = 0; i < N; i++)} $\rightarrow$ \texttt{for (i = 1; i < N; i++)} \\
          \hline
          FSM transition error & Redirects FSM next-state logic to an incorrect but valid state under certain conditions to disrupt intended flow. \\
          \hline
      \end{tabular}
    \end{center}
    \caption{Summary of mutation classes and their descriptions}
    \vspace{-12pt}
    \label{tab:mutationclasses}
\end{table}

\textbf{Single-line mutations}, such as bitwise corruption or logic bug, by definition, only apply to a single line of Verilog code. These parody nuanced, yet controlled, logical bugs that human engineers may introduce such as misused signals, breaks in logical conditions, etc. In spite of their limited scope of modification, these mutations still enable complex, functional reasoning. For example, the following wrong assignment mutation involves assigning data to the wrong variable. To apply this mutation correctly without triggering syntax failures, we must know the bit-width and purpose of nearby variables. In this case, \texttt{addr\_incr} is a single-bit signal derived from a logical condition. This implies that both \texttt{decr\_en\_i} and \texttt{incr\_en\_i} are also single-bit inputs, allowing them to be swapped:

\begin{lstlisting}[style=diffstyle]
assign addr_incr = (state_q == Checking) && (addr_q != LastAddr);
...
.incr_en_i(addr_incr),
@- .decr_en_i(1'b0), @-
@+ .incr_en_i(1'b0), @+
\end{lstlisting}

On the other hand, \textbf{multi-line mutations}, such as FSM transition errors and loop modifications apply to multiple lines of Verilog code. These mutations simulate more sophisticated faults that emerge across multiple functionally related lines---such as changing a loop's exist condition while simultaneously altering an internal control signal. This functionality requires a richer understanding of logical flow and control dependencies across regions, which current bug insertion solutions such as Certitude and Mantra lack as they rely on predefined syntactic alterations without functional awareness. 

For instance, the following loop modification mutation alters the starting index of the loop and modifies logic inside the loop body, causing a value to be ignored. Together, both changes simulate a plausible logical error.

\begin{lstlisting}[style=diffstyle]
@- for (i = 0; i < SIZE; i++) begin @-
@-   sum = sum + data[i]; @-
@+ for (i = 1; i < SIZE; i++) begin @+
@+   sum = sum + data[i-1]; @+
\end{lstlisting}

\section{LLM Output Examples}
\label{appendix:llmexamples}

This section provides concrete examples of outputs generated by the LLM agents at each step of the mutation pipeline. These examples were removed from the main body for brevity but are included here to illustrate the pipeline’s reasoning capabilities and mutation behavior.

\subsection*{Module Partition Example}

The following output from the module splitter illustrates how a large SystemVerilog module is divided into discrete, functionally meaningful regions:

\begin{llmoutput}
\textbf{Region 0 (Lines 11-80):} AES control FSM module declaration with interface ports \\
\textbf{Region 1 (Lines 82-84):} Unused input signal assignment \\
\textbf{Region 2 (Lines 86-90):} Tie-off signal generation for unused inputs \\
\textbf{Region 3 (Lines 92-100):} Signal declarations for the FSM and counters \\
\textbf{Region 4 (Lines 102-103):} Calculation for regular rounds in cipher operation \\
\textbf{Region 5 (Lines 105-448):} Always\_comb block defining FSM state transitions and outputs \\
\textbf{Region 6 (Lines 450-494):} Instantiation of Assertions related to FSM behaviors
\end{llmoutput}

\subsection*{Step 1: Region Selection}

The region selector agent then chooses a region and justify its decision:

\begin{llmoutput}
\textbf{Selected: Region 5}

Region 5 spans 344 lines and defines the FSM state transitions — a core behavioral block. It has seen fewer past mutations relative to its size, making it ideal for coverage. Given its role in logic flow, a \texttt{logic\_bug} mutation here may meaningfully disrupt output behavior and aid in testbench evaluation.
\end{llmoutput}

\subsection*{Step 2: Mutation Selection}

With a region selected, the mutation selector agent then selects a mutation class and location:

\begin{llmoutput}
\textbf{Chosen Mutation:} \texttt{logic\_bug} \\
\textbf{Target Line:} 370

Line 370 modifies the \texttt{sub\_bytes\_out\_ack\_o} signal conditioned on \texttt{dec\_key\_gen\_q\_i}. I will negate this condition to simulate a subtle logic fault. This location is strategic, as it's inside an if-else block whose behavior governs FSM handshaking signals.
\end{llmoutput}

\subsection*{Step 3: Mutation Insertion}

Finally, the mutation injector applies the edit and describes its effect:

\begin{llmoutput}
\textbf{Mutation Type:} \texttt{logic\_bug} \\
\textbf{Original Condition:} \texttt{if (out\_valid\_o \&\& out\_ready\_i)} \\
\textbf{Mutated Condition:} \texttt{if (out\_valid\_o || out\_ready\_i)}

By relaxing the condition from an AND to an OR, the block is more permissive. This could allow premature transitions or acknowledgment signaling, leading to observable downstream issues.
\end{llmoutput}

Each of these examples reflects the LLM agents’ ability to reason about structure, function, and mutation impact, enabling diverse and semantically plausible bug generation.

\section{Parallel Execution Strategy}
\label{appendix:parallelism}

A key advantage of our pipeline is its ability to run across multiple designs or multiple independent Verilog modules within the same design in parallel. We refer to these parallelized execution paradigms as \textbf{inter-design parallelism} and \textbf{intra-design parallelism} respectively.

The first, inter-design parallelism, aims to address the considerable inconsistencies in simulation times between different designs. If a verification engineer selects this option and specifies a set of multiple designs within the configuration file, the pipeline will automatically generate workflows for each and dispatch the appropriate threads to insert and evaluate bugs in parallel across all of them. Each of these threads is capable of writing to and reading from the shared mutation cache, allowing them to interactively learn from each other in a design-agnostic manner.

The second, intra-design parallelism is an extension of the first that follows a similar process, but dispatches a unique thread per specified SystemVerilog module. This is designed for verification engineers that have a particular design of interest which they aim to mutate and evaluate. However, if sufficient simulation resources are available, this paradigm is capable of operating across multiple designs as well, effectively flattening the entire workflow into a large set of modules.

\section{Pipeline Extensibility}

Our system is designed to be both modular and extensible. While the baseline pipeline itself has been demonstrated to adapt to unfamiliar contexts without changes, engineers are still provided high-level control over all of the following:

\begin{itemize}
    \item \textbf{Mutation classes}: Engineers can configure the mutation index and supply custom mutation specifications to allow new bug types and insertion strategies with minimal changes. They can optionally create multiple mutation indexes selected depending on the design and module.
    \item \textbf{Region definitions}: If there is a specific way modules should be segmented into regions, or if there are certain Verilog patterns or blocks that should be avoided entirely, engineers can easily specify that by revising the module splitter’s rulebook. 
    \item \textbf{Mutation target regions}: Engineers can optionally isolate specific design files and even specific mutation target regions within those files to constrain bug insertion to those particular areas. The pipeline will automatically accommodate coverage across those provided areas as demonstrated in the previous subsection.
    \item \textbf{Parallelism}: Depending on available computational resources, engineers can parallelize the workflow to expedite the overall rate of bug production (involving both generation and validation).
    \item \textbf{Bug scenario definition}: Engineers can also specify exactly what constitutes a given bug scenario. This includes the number of mutations per bug scenario, which mutations should and should not be inserted together, relative constraints on location between grouped mutations, and much more.
\end{itemize}

This extensible design ensures that the pipeline can scale to support customized, downstream verification tasks across different projects and environments.

\section{LLM Robustness and Reproducibility}
\label{appendix:llmrobustness}

As our pipeline relies on LLMs, which are inherently stochastic, we implement several measures to ensure consistency between different models, environments, and runs. These are detailed as follows:

\begin{itemize}
    \item \textbf{Mutation classes as guardrails:} Our structured mutation index constrains generation to predefined classes, each written and specified by a human engineer. This allows the pipeline to perform guided bug insertion, automatically producing complex mutations while limiting sources of variability.
    \item \textbf{Standardized prompts and in-context feedback:} We also utilize detailed and succinct prompt templates to guide the model toward valid and diverse mutation behavior across runs. These templates are injected with relevant subsets of the shared mutation cache and mutation index to limit hallucination and ensure that the pipeline can learn from its mistakes in context. Thus, any preliminary sensitivity to prompt details are quickly remedied by in-context learning.
    \item \textbf{Self-correction:} As detailed in the methods and results, all generated mutations are syntactically and functionally validated, with automatic rollback and retry logic. Thus, any variability in LLM architecture or target design will not compromise the end-to-end functionality or autonomy of the pipeline.
\end{itemize}

Ultimately, these techniques help ensure consistency while preserving the pipeline’s flexibility and autonomy.

\end{document}